\documentclass[11pt]{article}
\usepackage[russian]{babel}
\usepackage{amssymb,amsmath,amsthm}
\newcommand{\bc}{\begin{center}}
\newcommand{\ec}{\end{center}}
\newcommand{\be}{\begin{equation}}
\newcommand{\ee}{\end{equation}}
\newcommand{\bea}{\begin{eqnarray}}
\newcommand{\eea}{\end{eqnarray}}
\newcommand{\ba}{\begin{array}}
\newcommand{\ea}{\end{array}}
\newcommand{\edc}{\end{document}}

\def\l{\lambda}

\begin{document}
УДК 517.98
\begin{center}
\textbf{\Large {Трансляционно-инвариантные меры Гиббса для
плодородных HC-моделей с тремя состояниями на
дереве Кэли}}\\
\end{center}

\begin{center}
Р.М.Хакимов\footnote{Институт математики, ул. Дурмон йули, 29, Ташкент, 100125, Узбекистан.\\
E-mail: rustam-7102@rambler.ru}
\end{center}

\begin{abstract} В этой работе рассмотрены плодородные Hard-Core (HC) модели
с параметром активности $\lambda>0$ и тремя состояниями на дереве
Кэли порядка три. Известно, что существуют четыре типа таких
моделей, для двух из которых при $\lambda>0$
трансляционно-инвариантная мера Гиббса единственна, а для двух
других моделей найдена $\lambda_{cr}$ такая, что при
$\lambda>\lambda_{cr}$ существуют только три
трансляционно-инвариантные HC-меры Гиббса, а при
$\lambda\leq\lambda_{cr}$-только одна трансляционно-инвариантная
HC-мера Гиббса.
\end{abstract}

\textbf{Ключевые слова}: дерево Кэли, конфигурация, НС-модель,
мера Гиббса, трансляционно-инвариантные меры.

\section{Введение}\

Описание всех предельных мер Гиббса, в частности,
трансляционно-инвариантных мер Гиббса, для данного гамильтониана
является одним из основных задач теории гиббсовских мер.
Определение меры Гиббса и понятия, связанные с этой теорией,
вводятся стандартным образом (см. например \cite {6}-\cite {Si}).

Известно, что для модели Изинга на дереве Кэли количество
трансляционно-инвариантных гиббсовских мер, зависящихся от
температуры, может быть одно или три. В работе \cite{1} доказано,
что на дереве Кэли трансляционно-инвариантная мера Гиббса
антиферромагнитной модели Поттса с внешним полем единственна.
Работа \cite{Ga13} посвящена модели Поттса со счетным числом
состояний и c ненулевым внешним полем на дереве Кэли. Доказано,
что эта модель имеет единственную трансляционно-инвариантную меру
Гиббса.  В работе \cite {KRK} дано полное описание
трансляционно-инвариантных мер Гиббса для ферромагнитной модели
Поттса с $q-$состояниями и показано, что их количество равно
$2^q-1.$

В работе \cite{7} изучена HC (Hard Core)-модель на дереве Кэли и
доказано, что транцляционно-инвариантная мера Гиббса для этой
модели единственна. Кроме того, при некоторых условиях на
параметры НС-модели доказана неединственность периодических мер
Гиббса с периодом два. В работе \cite{XR} изучены слабо
периодические меры Гиббса для HC-модели. Показана единственность
(трансляционно-инвариантность) слабо периодической меры для
HC-модели.

В работах \cite{MRS},\cite{Ro} были изучены гиббсовские меры для
(hard core) HC-модели с тремя состояниями на дереве Кэли порядка
$k\geq1$. В работе \cite{bw} выделены плодородные HC модели,
соответствующие графам "петля"\,, "свисток"\,, "жезл"\, и "ключ".
В работе \cite{MRS} были изучены трансляционно-инвариантные и
периодические меры Гиббса для HC модели в случае "ключ"\, на
дереве Кэли и была доказана единственность
трансляционно-инвариантной меры Гиббса для любой положительной
активности $\lambda$. В \cite{Ro} были изучены
трансляционно-инвариантные и периодические меры Гиббса для HC
модели в случаях "свисток"\,, "жезл"\, и "петля"\, и показано, что
(случай "петля") из положительной активности не следует
единственность трансляционно-инвариантных мер.

В настоящей работе рассматриваются плодородные HC-модели с тремя
состояниями на однородном дереве Кэли. В случаях "петля" и "жезл"
описаны трансляционно-инвариантные HC-меры Гиббса на дереве Кэли
третьего порядка.

\section{Определения и известные факты}\

Дерево Кэли $\Im^k$ порядка $ k\geq 1 $ - бесконечное дерево, т.е.
граф без циклов, из каждой вершины которого выходит ровно $k+1$
ребер. Пусть $\Im^k=(V,L,i)$, где $V$ есть множество вершин
$\Im^k$, $L$ - его множество ребер и $i$ - функция инцидентности,
сопоставляющая каждому ребру $l\in L$ его концевые точки $x, y \in
V$. Если $i (l) = \{ x, y \} $, то $x$ и $y$ называются  {\it
ближайшими соседями вершины} и обозначаются через $l = \langle x,
y \rangle $. Расстояние $d(x,y), x, y \in V$ на дереве Кэли
определяется формулой
$$
d (x, y) = \min \ \{d | \exists x=x_0, x_1,\dots, x _ {d-1},
x_d=y\in V \ \ \mbox {такой, что} \ \ \langle x_0,
x_1\rangle,\dots, \langle x _ {d-1}, x_d\rangle\} .$$

Для фиксированого $x^0\in V$ обозначим $ W_n = \ \{x\in V\ \ | \ \
d (x, x^0) =n \}, $
\begin{equation}\label{p*}
 V_n = \ \{x\in V\ \ | \ \ d (x, x^0) \leq n \},\ \ L_n = \ \{l =
\langle x, y\rangle \in L \ \ | \ \ x, y \in V_n \}.
\end{equation}

Рассмотрим HC-модель ближайших соседей с тремя состояниями на
однородном дереве Кэли. В этой модели каждой вершине $x$ ставится
в соответствие одно из значений $\sigma (x)\in \{0,1,2\}$.
Значения $\sigma (x)=1,2$ означают, что вершина $x$ `занята', а
значение $\sigma (x)=0$ означает, что вершина $x$ `вакантна'.

Конфигурация $\sigma=\{\sigma(x),\ x\in V\}$ на дереве Кэли
задается как функция из $V$ в $\{0,1,2\}$. Множество всех
конфигураций на $V$ обозначается через $\Omega$. Аналогичном
образом можно определить конфигурации в $V_n$ ($W_n$), и множество
всех конфигураций в $V_n$ ($W_n$) обозначается как $\Omega_{V_n}$
($\Omega_{W_n}$).

Рассмотрим четыре типа плодородных (fertile) графов с тремя
вершинами $0,1,2$ (на множестве значений $\sigma(x)$), которые
имеют следующие виды:

\[
\begin{array}{ll}
\mbox{\it петля}: &  \{0,0\}\{0,1\}\{0,2\}\{1,1\}\{2,2\};\\
\mbox{\it жезл}: &  \{0,1\}\{0,2\}\{1,1\}\{2,2\};\\
\mbox{\it ключ}: &  \{0,0\}\{0,1\}\{0,2\}\{1,1\};\\
\mbox{\it свисток}: &  \{0,0\}\{0,1\}\{1,2\}.\\
\end{array} \]

Графы, которые не являются плодородными, называются бесплодными
(sterile) (см.\cite{bw}).

Пусть $O=\{\textit{ключ, жезл, петля, свисток}\}, \ G\in O.$
Конфигурация $\sigma$ называется $G$-\textit{допустимой
конфигурацией} на дереве Кэли (в $V_n$ или $W_n$), если $\{\sigma
(x),\sigma (y)\}$-ребро $G$ для любой ближайшей пары соседей $x,y$
из $V$ (из $V_n$). Обозначим множество $G$-допустимых конфигураций
через $\Omega^G$ ($\Omega_{V_n}^G$).

Множество активности \cite{bw} для графа $G$ есть функция $\l:G
\to R_+$ из множества $G$ во множество положительных дейсвительных
чисел. Значение $\l_i$ функции $\l$ в вершине $i\in\{0,1,2\}$
называется ее ``активностью''.

Будем писать $x<y,$ если путь от $x^0$ до $y$ проходит через $x$.
Вершина $y$ называется прямым потомком $x$, если $y>x$ и $x, \ y$
являются соседями. Через $S(x)$ обозначим множество прямых
потомков $x.$ Заметим, что в $\Im^k$ всякая вершина $x\neq x^0$
имеет $k$ прямых потомков, а вершина $x^0$ имеет $k+1$ потомков.

Для $\sigma_n\in\Omega_{V_n}^G$ положим
$$\#\sigma_n=\sum\limits_{x\in V_n}{\mathbf 1}(\sigma_n(x)\geq 1)$$
число занятых вершин в $\sigma_n$.

Пусть $z:\;x\mapsto z_x=(z_{0,x}, z_{1,x}, z_{2,x}) \in R^3_+$
векторнозначная функция на $V$. Для $n=1,2,\ldots$ и $\l>0$
рассмотрим вероятностную меру $\mu^{(n)}$ на $\Omega_{V_n}^G$,
определяемую как
\begin{equation}\label{rus2.1}
\mu^{(n)}(\sigma_n)=\frac{1}{Z_n}\lambda^{\#\sigma_n} \prod_{x\in
W_n}z_{\sigma(x),x}.
\end{equation}

Здесь $Z_n$-нормирующий делитель:
$$
Z_n=\sum_{{\widetilde\sigma}_n\in\Omega^H_{V_n}}
\lambda^{\#{\widetilde\sigma}_n}\prod_{x\in W_n}
z_{{\widetilde\sigma}(x),x}.
$$

Говорят, что вероятностная мера $\mu^{(n)}$ является
согласованной, если $\forall$ $n\geq 1$ и
$\sigma_{n-1}\in\Omega^G_{V_{n-1}}$:

\begin{equation}\label{rus2.2}
\sum_{\omega_n\in\Omega_{W_n}}
\mu^{(n)}(\sigma_{n-1}\vee\omega_n){\mathbf 1}(
\sigma_{n-1}\vee\omega_n\in\Omega^G_{V_n})=
\mu^{(n-1)}(\sigma_{n-1}).
\end{equation}
В этом случае существует единственная мера $\mu$ на $(\Omega^G,
\textbf{B})$ такая, что для всех $n$ и $\sigma_n\in
\Omega^G_{V_n}$
$$\mu(\{\sigma|_{V_n}=\sigma_n\})=\mu^{(n)}(\sigma_n).$$

\textbf{Определение.} Мера $\mu$, определенная формулой
(\ref{rus2.1}) с условием (\ref{rus2.2}), называется
($G$-)HC-\textit{мерой Гиббса} с $\lambda>0$,
\textit{соответствующей функции} $z:\,x\in V
\setminus\{x^0\}\mapsto z_x$. Множество таких мер (для
всевозможных $z$) обозначается через ${\mathcal S}_G$.

Пусть $L(G)$-множество ребер графа $G$, обозначим через $A\equiv
A^G=\big(a_{ij}\big)_{i,j=0,1,2}$ матрицу смежности $G$, т.е.
$$ a_{ij}\equiv a^G_{ij}=\left\{\begin{array}{ll}
1,\ \ \mbox{если}\ \ \{i,j\}\in L(G),\\
0, \ \ \mbox{если} \ \  \{i,j\}\notin L(G).
\end{array}\right.$$

В следующей теореме сформулировано условие на $z_x$, гарантирующее
согласованность меры $\mu^{(n)}$.

\textbf{Теорема 1.}\cite{Ro}\label{rust1} Вероятностные меры
$\mu^{(n)}$, $n=1,2,\ldots$, заданные формулой (\ref{rus2.1}),
согласованны тогда и только тогда, когда для любого $x\in V$ имеют
место следующие равенства:
\begin{equation}\label{rus2.3}\begin{array}{llllll}
z'_{1,x}=\lambda \prod_{y\in S(x)}{a_{10}+
a_{11}z'_{1,y}+a_{12}z'_{2,y}\over
a_{00}+a_{01}z'_{1,y}+a_{02}z'_{2,y}},\\[4mm]
z'_{2,x}=\lambda \prod_{y\in S(x)}{a_{20}+
a_{21}z'_{1,y}+a_{22}z'_{2,y}\over
a_{00}+a_{01}z'_{1,y}+a_{02}z'_{2,y}},
\end{array}
\end{equation}
где $z'_{i,x}=\lambda z_{i,x}/z_{0,x}, \ \ i=1,2$.\

\section{Трасляционно-инвариантные меры Гиббса}\

Мы полагаем, что $z_{0,x}\equiv 1$ и $z_{i,x}=z'_{i,x}>0,\ \
i=1,2$. Тогда для любых функций $x\in V\mapsto
z_x=(z_{1,x},z_{2,x})$, удовлетворяющих равенству

\begin{equation}\label{rus3.1}
z_{i,x}=\lambda \prod_{y\in S(x)}{a_{i0}+
a_{i1}z_{1,y}+a_{i2}z_{2,y}\over
a_{00}+a_{01}z_{1,y}+a_{02}z_{2,y}}, \ \ i=1,2,
\end{equation}
существует единственная $G$-HC-мера Гиббса $\mu$ и наоборот.
Начнем с трансляционно-инвариантных решений, в которых $z_x=z\in
R^2_+$, $x\neq x_0$.

\subsection{Случай $G=\textit{петля}$}\

В этом случае предполагая $z_x=z$, из (\ref{rus3.1}) получим
следующую систему уравнений:

\begin{equation}\label{rus3.2} \left\{\begin{array}{ll}
z_1=\lambda\left({1+ z_1\over 1+z_1+z_2}\right)^k,\\[2mm]
z_2=\lambda\left({1+z_2\over 1+z_1+z_2}\right)^k.
\end{array}\right.
\end{equation}
Вычитая из первого уравнения системы (\ref{rus3.2}) второе, имеем
$$
(z_1-z_2)\left[1-\lambda
\frac{(1+z_1)^{k-1}+...+(1+z_2)^{k-1}}{(1+z_1+z_2)^k}\right]=0.$$
Следовательно, $z_1=z_2$ или
\begin{equation}\label{rus3.3}
(1+z_1+z_2)^k=\lambda \big((1+z_1)^{k-1}+...+(1+z_2)^{k-1}\big),
\end{equation}
для $z_1\ne z_2.$ Для $z_1=z_2=z$ из системы (\ref{rus3.2}) имеем
\begin{equation}\label{rus3.4}
\lambda^{-1}z=f(z)=\left({1+z\over 1+2z}\right)^k.
\end{equation}

Функция $f(z)$ убывающая при $z>0$. Значит уравнение
(\ref{rus3.4}) имеет единственное решение $z^*=z^*(k, \lambda)$
для любой $\lambda>0.$

При $k=2$ известна следующая

\textbf{Теорема 2.}\cite{Ro} \textit{При $k=2$ для случая
$G=\textit{петля}$ верны следующие}
\begin{itemize}

\item[1)] \textit{При $\lambda\leq {9\over 4}$ существует единственная трансляционно-инвариантная мера Гиббса $\mu_0$ для Hard-Core модели;}

\item[2)] \textit{При $\lambda> {9\over 4}$ существуют не менее трех Hard-Core трансляционно-инвариантных  мер Гиббса $\mu_i,$ $i=0,1,2$.}
\end{itemize}

Следующее утверждение дает оценки для произвольного решения
системы уравнений (\ref{rus3.2}).

\textbf{Утверждение 1.} \textit{Решения системы уравнений
(\ref{rus3.2}) удовлетворяют неравенствам}
$${\lambda\over(1+\lambda)^k}<z_i<\lambda, \ i=1,2.$$\
\textbf{Доказательство.} Так как
$${1+ z_1\over 1+z_1+z_2}<1, \ {1+ z_2\over
1+z_1+z_2}<1,$$ то из (\ref{rus3.2}) получим, что $0<z_1<\lambda,
\ 0<z_2<\lambda.$ Кроме того, используя эти неравенства, из
первого уравнения (\ref{rus3.2}) имеем
$${\lambda\over z_1}=\left({1+ z_1+z_2\over 1+z_1}\right)^k=
\left(1+{z_2\over 1+z_1}\right)^k<\left(1+{\lambda\over
1+0}\right)^k=(1+\lambda)^k \ \Rightarrow
z_1>{\lambda\over(1+\lambda)^k}.$$ Из второго уравнения
(\ref{rus3.2}), аналогично, можно получить
$$z_2>{\lambda\over(1+\lambda)^k}.$$
Утверждение доказано.

\textbf{Случай $k=3$.} В этом случае система уравнений
(\ref{rus3.2}) при $\sqrt[3]{z_1}=x>0, \ \sqrt[3]{z_2}=y>0$ имеет
вид

\begin{equation}\label{rus3.22} \left\{\begin{array}{ll}
x=\sqrt[3]{\lambda}{1+ x^3\over 1+x^3+y^3},\\[3 mm]
y=\sqrt[3]{\lambda}{1+ y^3\over 1+x^3+y^3}.
\end{array}\right.
\end{equation}
Разделив первое уравнение на второе в последней системе уравнений,
получим
$$(x-y)\left(x+y-{1\over xy}\right)=0.$$
Отсюда $x=y$ или
\begin{equation}\label{rus3.23}
x+y={1\over xy}
\end{equation}
при $x\neq y$.

Из уравнения (\ref{rus3.23}) при $x\neq y$ найдем $y$:
\begin{equation}\label{rus3.24}
y={\sqrt{x^4+4x}-x^2\over 2x}.
\end{equation}

Далее, из (\ref{rus3.23}), заметив $y^2={1\over x}-xy$, перепишем
первое уранение системы (\ref{rus3.22}) следующим образом:
$$x=\sqrt[3]{\lambda}{1+ x^3\over 1+x^3+y^3}=\sqrt[3]{\lambda}{1+ x^3\over 1+x^3+y\cdot y^2}=
\sqrt[3]{\lambda}{1+ x^3\over 1+x^3+y\cdot ({1\over x}-xy)}=$$
$$=\sqrt[3]{\lambda}{1+ x^3\over 1+x^3+{y\over x}-xy^2}=\sqrt[3]{\lambda}{1+ x^3\over 1+x^3+{y\over x}-x({1\over x}-xy)}=
\sqrt[3]{\lambda}{(1+ x^3)x\over x^4+y(1+x^3)}.$$ Следовательно,
используя (\ref{rus3.24}), будем иметь уравнение
$$ x^4+y(1+x^3)=\sqrt[3]{\lambda}(1+ x^3) \ \Rightarrow \  y={\sqrt[3]{\lambda}(1+ x^3)-x^4\over 1+x^3} \ \Leftrightarrow  \ {\sqrt{x^4+4x}-x^2\over 2x}={\sqrt[3]{\lambda}(1+ x^3)-x^4\over 1+x^3},$$
которое эквивалентно уравнению
$$4x(\sqrt[3]{\lambda}x^8-\sqrt[3]{\lambda^2}x^7+2x^6-2\sqrt[3]{\lambda^2}x^4+2x^3-\sqrt[3]{\lambda}x^2-\sqrt[3]{\lambda^2}x+1)=0.$$
Это уравнение имеет решение $x=x(\lambda).$ Но мы рассмотрим это
уравнение относительно переменной $\lambda$ и получим решение
$\lambda=\lambda(x).$ Для этого, обозначив $\sqrt[3]{\lambda}=t$,
рассмотрим
\begin{equation}\label{rus3.24}
x(x^3+1)^2t^2-x^2(x^6-1)t-(2x^6+2x^3+1)=0.
\end{equation}
Отсюда
$$t_{1,2}={x^2(x^3-1)\pm (x^3+1)\sqrt{x^4+4x}\over 2x(x^3+1)},$$
т.е. уравнение (\ref{rus3.24}) приводится к виду
$$x(x^3+1)^2(t-t_1)(t-t_2)=0,$$
где
$$t_{1}={x^2(x^3-1)+ (x^3+1)\sqrt{x^4+4x}\over 2x(x^3+1)};$$
$$t_{2}={x^2(x^3-1)- (x^3+1)\sqrt{x^4+4x}\over 2x(x^3+1)}.$$
Следовательно, т.к. $t_2<0,$ то
$$t-t_1=0 \ \Leftrightarrow  t={x^2(x^3-1)+ (x^3+1)\sqrt{x^4+4x}\over 2x(x^3+1)}=\varphi(x).$$
Проанализируя функцию $\varphi(x),$ заметим, что $\varphi(x)>0.$
Кроме того, т.к. $\varphi(x)\rightarrow +\infty$ при $x\rightarrow
0$ и $x\rightarrow+\infty$, то каждому значению $t$ соответствует
по крайней мере два значения $x$ при $t>\varphi(x^{*})$, одно
значение при $t=\varphi(x^{*})$ и уравнение $t=\varphi(x)$ не
имеет решений при $t<\varphi(x^{*}),$ где $x^{*}$ есть решение
уравнения $\varphi'(x)=0.$ С помощью программы MathCAD (Maple)
найдем положительное решение уравнения $\varphi'(x)=0,$ которое
единственно и равно
$$x^{*}={\sqrt[3]{4}\over 2}.$$
Отсюда
$$\varphi(x^{*})={2\sqrt[3]{4}\over 3}$$ и
$$\lambda_{cr}=t^3=\varphi^3(x^{*})={32\over 27}.$$

Заметим, что если $\varphi''(x)>0$, то каждому значению $\lambda$
соответствует только два значения $x$ при $\lambda>\lambda_{cr}$.
Поэтому докажем, что $\varphi''(x)>0$. Действительно,
$$\varphi''(x)={12x^4(2-x^3)(x+4)\sqrt{x^2+4x}+(x^4+4x+12)(x^3+1)^3\over 2x^2(x+4)(x^3+1)^3\sqrt{x^2+4x}}.$$
Рассмотрим
$$\psi(x)=12x^4(2-x^3)(x+4)\sqrt{x^2+4x}+(x^4+4x+12)(x^3+1)^3.$$
Ясно, что если $x^3<2$, то $\varphi''(x)>0.$ Пусть $x^3>2.$ Тогда
из $\psi(x)>0$ получим
$$\alpha(x)=x^{26}+14x^{23}+24x^{22}+79x^{20}+240x^{19}-1492x^{17}-5976x^{16}-7776x^{15}+7327x^{14}+$$
$$+29568x^{13}+38448x^{12}-6466x^{11}-25368x^{10}-33984x^9+289x^8+1584x^7+$$
$$+2160x^6+104x^5+600x^4+864x^3+16x^2+96x+144>0.$$
Здесь $\alpha(1)=496>0.$ С помощью программы MathCAD (Maple) можно
увидеть, что уравнение $\alpha(x)=0$ имеет два положительных
решения $x_1<1, \  \ x_2<1.$ Отсюда получим, что $\varphi''(x)>0$
при $x>0$.

Пусть теперь  в (\ref{rus3.23}) $x=y.$ Тогда $x=y={1\over
\sqrt[3]{2}}.$ Используя первое или второе уравнение системы
(\ref{rus3.22}), найдем $\lambda={32\over 27}.$ Отсюда, заметим,
что при этом значении $\lambda$ значения $z=z_1=z_2={1\over 2}$
удовлетворяют уравнению (\ref{rus3.4}), т.е. $({1\over 2}, {1\over
2})$ является единственным решением системы (\ref{rus3.2}).\

Из всего выше сказанного верна следующая

\textbf{Теорема 3.} \textit{При $k=3$ и $\lambda_{cr}={32\over
27}$ для случая $G=\textit{петля}$ верны следующие утверждения:}

\textit{1. При $\lambda>\lambda_{cr}$ существуют только три
трансляционно-инвариантные меры Гиббса для HC-модели.}

\textit{2. При $\lambda\leq\lambda_{cr}$ существует только одна
трансляционно-инвариантная мера Гиббса для HC-модели.} \

\subsection{Случай $G=\textit{жезл}$}\

В этом случае предполагая $z_x=z$, из (\ref{rus3.1}) при $k=3$
получим следующую систему уравнений:

\begin{equation}\label{rus3.26} \left\{\begin{array}{ll}
z_1=\lambda\left({1+ z_1\over z_1+z_2}\right)^3,\\[2mm]
z_2=\lambda\left({1+z_2\over z_1+z_2}\right)^3
\end{array}\right.
\end{equation}
или при $\sqrt[3]{z_1}=x>0, \ \sqrt[3]{z_2}=y>0$ имеет вид

\begin{equation}\label{rus3.27} \left\{\begin{array}{ll}
x=\sqrt[3]{\lambda}{1+ x^3\over x^3+y^3},\\[3 mm]
y=\sqrt[3]{\lambda}{1+ y^3\over x^3+y^3}.
\end{array}\right.
\end{equation}

Разделив первое уравнение на второе в последней системе уравнений,
получим
$$(x-y)\left(x+y-{1\over xy}\right)=0.$$
Отсюда $x=y$ или
\begin{equation}\label{rus3.29}
x+y={1\over xy}
\end{equation}
при $x\neq y$.

Отсюда следует два случая: 1. $x<1, \ y<1 \ (\Rightarrow
\lambda<1)$ или 2. $x<1(x>1), \ y>1(y<1).$

Следующее утверждение дает оценки для произвольного решения
системы уравнений (\ref{rus3.26}).

\textbf{Утверждение 2.} \textit{Решения системы уравнений
(\ref{rus3.26}) удовлетворяют неравенствам}

1. При $z_1<1, \ z_2<1$ верно
$$\lambda<z_i<{(1+\lambda)^3\over8\lambda^2}, \ i=1,2.$$

2. При $z_1<1(z_1>1), \ z_2>1(z_2<1)$ верно
$$0<z_1<\lambda, \ \lambda<z_2<{(1+\lambda)^3\over \lambda^2}.$$ \

\textbf{Доказательство.} 1. Так как $z_1<1, \ z_2<1$, то
$z_1>\lambda, \ z_2>\lambda.$ Отсюда
$$ z_1=\lambda\left({1+z_1\over z_1+z_2}\right)^3<\lambda({1+z_1\over z_1+\lambda})^3=\psi(z_1).$$
Здесь производная $\psi'(z_1)<0,$ т.е. функция $\psi(z_1)$ убывает
при $z_1>\lambda.$ Следовательно,
$$\lambda <z_1<\lambda\left({1+\lambda \over 2\lambda}\right)^3={(1+\lambda)^3\over 8\lambda^2} \ \left(\Rightarrow \ \sqrt[3]{\lambda}<x<{1+\lambda\over 2\sqrt[3]{\lambda^2}}\right).$$
Аналогично,
$$\lambda <z_2<{(1+\lambda)^3\over 8\lambda^2} \ \left(\Rightarrow \ \sqrt[3]{\lambda}<y<{1+\lambda\over 2\sqrt[3]{\lambda^2}}\right).$$
2. Пусть  $z_1<1, \ z_2>1$. Тогда
$$z_1<1 \ \Rightarrow \ z_2=\lambda\left({1+z_2\over z_1+z_2}\right)^3>\lambda;$$
$$z_2>1 \ \Rightarrow \ z_1=\lambda\left({1+z_1\over z_1+z_2}\right)^3<\lambda,$$
т.е. $z_1<\lambda<z_2.$ Отсюда $0<z_1<\lambda \
(0<x<\sqrt[3]{\lambda})$ и
$$\lambda<z_2=\lambda\left({1+z_2\over z_1+z_2}\right)^3<\lambda\left({1+z_2\over z_2}\right)^3=\varphi(z_2).$$
Здесь производная $\varphi'(z_2)<0,$ т.е. функция $\varphi(z_2)$
убывает при $z_2>\lambda$. Следовательно,
$$\lambda<z_2<\lambda\left({1+\lambda\over \lambda}\right)^3={(1+\lambda)^3\over \lambda^2} \
\left(\Rightarrow \ \sqrt[3]{\lambda}<y<{1+\lambda\over
\sqrt[3]{\lambda^2}}\right).$$ Утверждение доказано.

Подобно случаю $G=\textit{петля}$ тем же методом можно получить
следующую теорему.

\textbf{Теорема 4.} \textit{При $k=3$ и $\lambda_{cr}={4\over 27}$
для случая $G=\textit{жезл}$ верны следующие утверждения:}

\textit{1. При $\lambda>\lambda_{cr}$ существуют только три
трансляционно-инвариантные меры Гиббса для HC-модели.}

\textit{2. При $\lambda\leq\lambda_{cr}$ существует только одна
трансляционно-инвариантная мера Гиббса для HC-модели.} \

\textbf{Замечание.} Заметим, что и в этом случае при
$\lambda\leq\lambda_{cr}$ единственной трансляционно-инвариантной
мере соответствует $z=z_1=z_2={1\over 2}.$ Кроме того, известно,
что при $k=2$ в случае $G=\textit{жезл}$ $\lambda_{cr}=1$ и
справедливо утверждение, подобное теореме 4 (\cite{Ro}).

\subsection{Случаи $G=\textit{свисток или ключ}$}\

В случаях $G=\textit{свисток или ключ}$, предполагая $z_x=z$, из
(\ref{rus3.1}) при любом $k$ имеем системы уравнений

$$ \left\{%
\begin{array}{ll}
z_1=\lambda\left({1+ z_2\over 1+z_1}\right)^k,\\[2mm]
z_2=\lambda\left({z_1\over 1+z_1}\right)^k,
\end{array}%
\right. $$

$$\left\{%
\begin{array}{ll}
z_1=\lambda\left({1+ z_1\over 1+z_1+z_2}\right)^k,\\[2mm]
z_2=\lambda\left({1\over 1+z_1+z_2}\right)^k,
\end{array}%
\right.$$
 соответственно.

В каждом из этих случаев было доказано, что  для любых $\lambda
>0$ и $k\geq 1$ трансляционно-инвариантная мера Гиббса
единственна (см.\cite{MRS}, \cite{Ro}).\

\textbf{Благодарность.} Автор выражает глубокую признательность
профессору У. А. Розикову за постановку задачи и полезные советы
по работе.

\end{document}